\documentclass[preprint,preprintnumbers,amsmath,amssymb,12pt]{revtex4-2}
\usepackage[colorlinks,linkcolor=red,citecolor=blue]{hyperref}
\usepackage{epsf,epsfig}
\usepackage{bm}
\usepackage{graphicx}
\usepackage{natbib}
\usepackage{color}
\usepackage{mathrsfs}
\usepackage{float}
\usepackage{multirow}
\usepackage{makecell}
\usepackage{booktabs}

\def\be{\begin{equation}}
\def\ee{\end{equation}}
\def\bea{\begin{eqnarray}}
\def\eea{\end{eqnarray}}

\begin{document}

\today\\

\title{Strong gravitational lensing by compact objects with magnetic dipole}

\author{Jin Young Kim \footnote{E-mail address: jykim@kunsan.ac.kr} }
\affiliation{Department of Physics, Kunsan National University,
Kunsan 54150, Korea}

\begin{abstract}

Compact objects with magnetic dipole are considered as gravitational lenses. The presence of strong magnetic field near the photon sphere can affect the trajectory of light. 
We compute the deflection angle near the photon sphere on the equatorial plane of the magnetic dipole. In the Einstein-Maxwell gravity we use the asymptotic metric when a massive object has a magnetic dipole moment.
In the Einstein-Born-Infeld gravity we use the effective metric accomodating the nonlinear electrodynamic effect. The deflection angles are expressed as perturbations from the Schwarzschild lensing. 
As a numerical estimate, we apply the result to a magnetar. 
\end{abstract}


\keywords{strong gravitational lensing, magnetic dipole, nonlinear electrodynamics }

\maketitle

\newpage

\section{Introduction}

Gravitational lensing based on general relativity is one of the important tools in astronomy and astrophysics. The deflection of light by massive object was mainly studied through the weak field approximation. After the discovery of the shadow image of black hole \cite{EHT1, EHT2, EHT3, EHT4}, the gravitational lensing in the strong field limit also has become an important issue. 
Virbhadra and Ellis \cite{virell} showed that, by strong gravitational lensing due to Schwarzschild black hole, a sequence of images can be formed for a light ray passing close to the photon sphere. Because any physical quantity that affects the energy-momentum tensor can cause the bending of light, strong gravitational lensing by compact objects other than the Schwarzschild black hole can also be considered. The formalism of strong gravitational lensing can be applied to Reissner-Nordstrom, Kerr and Kerr-Newman black holes. Bozza \cite{Bozza} estabilished a general formalism to compute the deflection angle for a compact object whose metric is spherically symmetric and static. He also studied strong gravitational lensing  on the equatorial plane of Kerr black hole \cite{bozza2}. 

It seems improbable to observe light bending by strong electric field because the observed universe is charge neutral and no strongly charged black hole has ever been observed. However one can think of the light bending by strong magnetic field because the magnetic field on the surface of a magnetar is estimated up to the order of $10^{11} {\rm T}$ \cite{duncan,thompson}. This field strength is the estimated lower bound of the Born-Infeld parameter \cite{Jackson} where nonlinear electrodynamic effects can be considered. In Einstein-Maxwell theory, the null geodesic of the electromagnetic wave is the same as the null geodesic of the gravitational wave. However, in Einstein-Born-Infeld theory, the null geodesic of the electromagnetic wave is not the same as that of the gravitational wave due to the nonlinear electrodynamic effects \cite{Plebanski}. 

In the previous work \cite{kim2022, kim2024}, we studied weak bending of light by a compact object with magnetic dipole using the effective null geodesic accomodating the nonlinear electrodynamic effects. Here, we consider the strong deflection of light when a light ray is passing close to the photon sphere. 
The purpose of this study is to find an analytic formula to see the dependence of the strong deflection angle on the parameters of the system. We confine our analysis on the equatorial plane of the magnetic dipole for simplicity. In this case, analytical treatment is available. 
In general the bending of light by compact objects with magnetic dipole depends on the relative orientation between the dipole axis and the incoming light ray. 

The organization of the paper is as follows. 
In Sec. II, we consider the geometrical aspects of a compact astrophysical object with a magnetic dipole moment in the Einstein-Maxwell gravity. Using the geodesic equation on the equatorial plane, we compute the deflection angle as a function of magnetic dipole. As a numerical example we apply the result to a magnetar and estimate the order of correction.
In Sec. III, we obtain the effective metric on the equatorial plane of magnetic dipole in Einstein-Born-Infeld theory. We compute the deflection angle as a perturbation up to leading order in Born-Infeld parameter. We also apply the result to a magnetar for order-of-magnitude estimation. 
In Sec. IV, we summarize and discuss.

\section{Deflection angle in Einstein-Maxwell gravity}

We are interested in the path of light close to the photon sphere of a compact object with magnetic fields. For example, one might consider a black hole having a magnetic dipole or a magnetar compact enough to have a photon sphere. 
Here we consider the strong defelction in the Einstein-Maxwell theory. When the compact object has a magnetic dipole, it is convenient to use the cylindrical coordinates, $x^\mu = ( t, r, z , \phi)$, taking the direction of magnetic dipole as $z$-axis. 
For the coupled Einstein-Maxwell equations to be well defined, the metric tensor and the electromagnetic four-potential can be written as
\be
g_{\mu \nu} =  \begin{pmatrix}  e^\rho  & 0  & 0  & 0\cr 
0 & -e^\lambda  &  0 & 0 \cr  
0&  0& -e^\lambda &0  \cr
0&  0& 0  & - r^2 e^{-\nu}  \end{pmatrix} ,  \label{metricftn}
\ee
\be
A_\mu = ( 0, 0, 0, - \psi ) , \label{four-pot}
\ee
where $\rho, \lambda, \nu$, and $\psi$  are functions of $r$ and $z$ only. Finding the exact solutions from the equations of motion is not easy.
However asymptotic solutions were obtained as expansions in powers of gravitational constant by Martin and Prechett \cite{martinpritchett}.
Up to the first order in $G$, the solutions are given by 
\bea
e^\rho &=& 1 - \frac {2GM} {v} + \frac {G \mu^2 z^2} {v^6} ,    \label{erho}   \\ 
e^\lambda &=& 1 + \frac {2GM} {v} - \frac {G \mu^2 (r^4  - 6 r^2 z^2 + 2 z^4 )} {2 v^8} ,  
                    \label{elambda} \\
e^{- \nu} &=& 1 + \frac {2GM} {v} - \frac {G \mu^2 z^2} { v^6 },      \label{enu} \\
\psi &=& \frac { \mu r^2} {v^3} \left ( 1 + \frac {GM} {2 v} \right ),   \label{psi} 
\eea
where $v = \sqrt{r^2 + z^2}$ is the spherical distance and $\mu$ is the magnetic dipole moment. 

On the equatorial plane, $z=0$, the metric can be written as 
\be 
ds^2 = \left ( 1 - \frac{ 2GM}{r} \right ) dt^2 -   \left( 1 + \frac{ 2GM} {r} - \frac {G \mu^2}{2 r^4} \right ) dr^2 - r^2 \left ( 1 + \frac{2 GM } {r} \right ) d \phi^2 .
\ee
To compute the deflection with the standard form of the metric, we define 
\be 
x = r \sqrt { 1 + \frac{2 GM } {r} } = r  \left ( 1 +  \frac{GM } {r}  + \cdots \right )  , 
\ee
or 
\be 
r = x \left (1 -  \frac{GM } {x}  + \cdots \right ) . 
\ee
In terms of $x$, we can write the effective metric to the leading order as
\be 
ds^2 =  A(x)dt^2 -  B(x) d x^2  -  C(x) d \phi^2 , \label{standardformABC}
\ee
where
\be 
A(x) = \left ( 1 - \frac{ 2GM}{x} \right ),  ~B(x) = 1 + \frac{ 2GM} {x} - \frac {G \mu^2}{2 x^4} ,~ C(x) =x^2  .
\ee

From the geodesic equation, for a light ray coming from infinity and going to infinity, the deflection angle can be obtained as a function of the distance of closest approach $x_0$ \cite{virell, Weinberg}
\bea
\varphi (x_0 ) &=& I (x_0 ) - \pi   \label{deflangle} ,  \\
I(x_0 ) &=& \int_{x_0 }^\infty \frac{2 \sqrt{B} }{  \sqrt{C} \sqrt { \frac{CA_0}{C_0 A} -1} } dx .   \label{deflang}
\eea
We compute the strong deflection angle following the formalism and notation proposed by Bozza \cite{Bozza}. Defining new variables
\bea
y &=&A (x ) ,    \\
z &=&  \frac{ y-y_0 }{  1 - y_0}  ,
\eea
where $y_0 = A_0$, the intergral in (\ref{deflang}) can be written as
\be 
I(x_0 ) = \int_{0 }^1 R(z, x_0 ) f(z, x_0 ) dz , \label{Ix0}
\ee
\be 
 R(z, x_0 )= \frac{ 2 \sqrt{B y } }{C A^\prime} (1 - y_0 ) \sqrt{C_0} ,  \label{rzx0}
\ee
\be 
 f(z, x_0 )= \frac{ 1 } {\sqrt {y_0 - [(1 - y_0 ) z + y_0 ]  \frac{C_0 }{C}   } } .   \label{fzx0}
\ee
The function $f(z,x_0)$ is divergent for $z \rightarrow 0$ and the order of divergence can be analyzed by expanding the argument of the square root to the second order in $z$
\be 
 f(z, x_0 ) \sim f_0 (z, x_0 )= \frac{ 1 } {\sqrt {\alpha z + \beta z^2  } } , 
\ee
where
\bea 
\alpha &=& \frac{ 1- y_0} {C_0 A_0^\prime } ( C_0^\prime y_0 - C_0 A_0^\prime ),  \label{alpha} \\
\beta &=& \frac{( 1- y_0 )^2} {2C_0^2 { A_0^\prime}^3 } [ 2 C_0  C_0^\prime {A_0^\prime}^2 
+ (  C_0  C_0^{\prime \prime} - 2  {C_0^\prime}^2 ) y_0  A_0^\prime -  C_0  C_0^\prime y_0 A_0^{\prime \prime}] . \label{beta}
\eea

Defining the Schwarzschild radius $x_s = 2GM=1$ as the measure of distances, we work with the scaled metric to the leading order
\be 
A(x) = \left ( 1 - \frac{1}{x} \right ),  ~B(x) = 1 + \frac{1} {x} - \frac {\nu^2}{2 x^4} ,~ C(x) =x^2  , \label{explicitABC}
\ee
where 
\be
\nu^2 =\frac {G \mu^2}{(2GM)^4} .    \label{defnu}
\ee  
The radius of the photon sphere, obtained from $\alpha =0$,
\be
\frac{ C^\prime (x)}  { C (x)} = \frac{ A^\prime (x)}  { A (x)},  \label{photonsphere}
\ee
is $ x_m = 3 /2$. Inserting this into Eq. (\ref{beta}), we obtain $\beta_m =1 $.
Both $x_m$ and $\beta_m$ are the same as those for the Schwarzschild black hole.

To the leading order the two functions in Eqs. (\ref{rzx0}) and  (\ref{fzx0}) can be written in terms of $z$ as
\be 
R(z, x_0 )= 2 \left [ 1 - \frac{ \nu^2} { 2}   \frac{1} { 1 - \frac{1-z}{x_0}} \frac { (1-z)^4} { x_0^4} + \cdots \right ] ,  \label{rzx0leading}
\ee
\be 
 f(z, x_0 )= \frac{ 1 } {\sqrt { (2 - \frac{3}{x_0} ) z + ( -1  + \frac{3}{x_0} ) z^2 - \frac{1}{x_0} z^3 } } .   \label{fzx0explicit}
\ee
The integral in Eq. (\ref{Ix0}) can be split into two parts
\be 
I(x_0 ) = I_D (x_0 ) + I_R (x_0 )  , \label{Ix0twoterms}
\ee
where 
\be 
I_D (x_0 ) = \int_{0 }^1 R(0, x_m ) f_0(z, x_0 ) dz , \label{IDx0}
\ee 
\be 
I_R (x_0 ) = \int_{0 }^1 g(z, x_0) dz , \label{IRx0}
\ee
\be 
g(z, x_0 ) = R(z, x_0 ) f(z, x_0 ) - R(0 , x_m ) f_0 (z, x_0 ) . \label{gzx0}
\ee

The divergent part of the integral $I_D (x_0 ) $ is solved exactly as
\be 
I_D (x_0 ) = R(0, x_m ) \frac{2} {\sqrt{\beta}} \ln \frac{\sqrt{\beta} + \sqrt{ \alpha + \beta}} {\sqrt{\alpha}}.  \label{IDx0int}
\ee 
Up to $O (x_0 - x_m )$, one can expand $\alpha$ as 
\be 
\alpha = \frac{ 2 \beta_m A_m^\prime }{1 - y_m } (x_0 -x_m ) + O(x_0 -x_m )^2 , \label{alpham}
\ee
where 
\be 
\beta_m = \beta {\big |}_{x_0 = x_m } 
= \frac { C_m (1 - y_m )^2 ( C_m^{\prime \prime} y_m - C_m A^{\prime \prime } (x_m ) ) } {2 y_m^2 {C_m^\prime}^2 } .  \label{betam}
\ee
Substituting Eqs. (\ref{alpham}) and  (\ref{betam}) into Eq. (\ref{IDx0int}), we have 
\be 
I_D (x_0 ) = -a \ln \left ( \frac{x_0}{x_m} - 1 \right ) + b_D + O (x_0 - x_m),  \label{IDx0ln+bD}
\ee
\be 
a = \frac{ R( 0, x_m )}{\sqrt {\beta_m}},    \label{aRbeta}   
\ee
\be 
b_D = \frac{ R( 0, x_m )}{\sqrt {\beta_m}} \ln \frac{ 2 ( 1 - y_m )}{A_m^\prime x_m } .    \label{bDRbeta}   
\ee
The regular part of the integral $I_R (x_0 )$ can also be expanded in powers of $(x_0 - x_m ) $
\be 
I_R (x_0 ) = \sum_{n=0}^{\infty} \frac{1}{n!} (x_0 - x_m )^n  \int_{0 }^1  \frac {\partial^n g } { \partial x_0^n } {\bigg |}_{x_0 = x_m } dz .
\label{Ix0Rpower}
\ee
Up to $O (x_0 - x_m ) $, we keep only the $n=0$ term. Then the contribution to the deflection angle by the regular part is 
\be 
b_R = I_R (x_m ). \label{bR}
\ee
From Eq. (\ref{deflangle}), the deflection angle close to the divergence can be written as 
\be
\varphi (x_0 ) = -a \ln \left ( \frac{x_0}{x_m} - 1 \right ) + b + O(x_0 -x_m ) , \label{varphi}
\ee
\be
b = - \pi +b_D + b_R . \label{bbdbr}
\ee
 
Because the distance of closest approach $x_0$ is coordinate dependent, we need to express the deflection angle in Eq. (\ref{varphi}) in terms of the impact parameter, which is coordinate independent and physically measurable.
From the conservation of angular momentum, the distance of closest approach is related to the impact parameter $u$ by
\be 
u =\sqrt{ \frac{C_0 } {A_0 }}.  \label{impactpara}
\ee
We can express Eq. (\ref{varphi}) in terms the angular separation $\theta = u/D_{OL}$, where $D_{OL}$ is the distance from the lens to the observer. 
Expanding Eq. (\ref{impactpara}) in terms of the the minimum impact parameter $ u_m =\sqrt{ {C_m } /{y_m }}$, we have
\bea
u - u_m &=& c ( x_0 - x_m )^2 ,  \\
c &=& \beta_m \sqrt { \frac{y_m } {C_m^3} } \frac{ {C_m^\prime}^2 }{ 2 (1 - y_m )^2 } .
\eea
Then we can express the deflection angle as a function of $\theta$ as 
\bea
\varphi (\theta)  &=& - {\bar a } \ln \left ( \frac{ \theta D_{OL}} {u_m }  -1 \right ) + {\bar b}  , \label{varphitheta}  \\
\bar a   &=&  \frac{a}{2} =  \frac{ R( 0, x_m )}{2 \sqrt {\beta_m}}   ,  \label{abar}   \\
\bar b   &=&  b + \frac{a}{2} \ln \frac{ c x_m^2} {u_m } = - \pi + b_R + \bar a \ln \frac{ 2 \beta_m}{y_m} . \label{bbar}
\eea

Using the explicit form of the metric in Eq. (\ref{explicitABC}), to the leading order, we obtain  
\be
\bar a =  1 -  \left ( \frac{2}{3} \right )^3 \nu^2 , \label{abarfinal}
\ee
\be 
b_R = b_{R,{\rm Sch}} - c_2 \left ( \frac{2}{3} \right )^4 \nu^2 ,   \label{bRfinal}
\ee
\be
u_m = \frac {3 \sqrt{3}}{2} ,
\ee
where  $b_{R,{\rm Sch}} = 2 \ln[ 6(2-\sqrt{3} )]$ is the regular term for the Schwarzschild black hole and $c_2$ is given by the integral
\be 
c_2 = \int_0^1 3 \left [ \frac{ (1-z)^4}{1 +2 z } - 1 \right ] \frac{dz } {\sqrt{ z^2 - \frac{2}{3} z^3 }} = -8.4407. 
\ee 
Inserting Eqs. (\ref{abarfinal}) and (\ref{bRfinal}) in Eq. (\ref{bbar}), we have
\be 
\bar b = - \pi + 2 \ln [ 6 (2 - \sqrt{3}) ] + \ln 6 - \nu^2 \left [ c_2 \left (  \frac{2}{3} \right )^4 + \left (  \frac{2}{3} \right )^3 \ln6  \right ] = {\bar b}_{\rm Sch} +1.1364 \nu^2 ,
\ee
where $ {\bar b}_{\rm Sch} = - \pi + \ln [ 216 (7 - 4\sqrt{3}) ] $. 
Finally we can write the deflection angle as
\be 
\varphi (\theta ) = - \left [1 - \left( \frac{2}{3} \right )^3 \nu^2  \right ] \ln \left ( \frac{ \theta D_{OL} } { u_m } -1 \right ) + \ln [216 ( 7 - 4 \sqrt{3} ) ]  - \pi +1.1364 \nu^2  . \label{varphiEM}
\ee
For a given nonzero value of $\mu$, the strong deflection coefficient $\bar a$ is smaller and $\bar b$ is larger than their Schwarzschild counterparts. 
Because the logathmic term dominates in Eq. (\ref{varphiEM}) for $u \sim u_m$, the overall effect of the magnetic dipole is decreasing the deflection angle. 

Now we consider some numerical estimation for hypothetical compact objects. To apply the result, the lensing object should be compact enough to have a photon sphere.
One can easily think of a black hole with strong magnetic field. However, it seems unprobable for a black holes to have a magnetic dipole by no-hair theorem. Next one can consider a magnetar having a photon sphere. 
For the typical neutron star with mass $1.4 M_{\rm sun}$, the Schwarzschild radius is $r_{\rm Sch} = 4.15 \rm km$. The radius of neutron star with this mass is estimated about $10 \rm km$. So we cannot think of strong bending because the photon sphere, $r_{\rm ps} = 1.5 r_{\rm Sch} = 6.22 \rm km$, is inside the neutron star. Nontheless, assuming more compact magnetars (compact enough to have the photon sphere outside) might exist, we try the order-of-magnitude estimation. 

Restoring the simplified units in Eq. (\ref{defnu}), we have
\be
\nu^2 = \frac {G } {c^4} \frac{ \mu_0}{4 \pi} \frac{\mu^2} { (2 GM/c^2 )^4} .  \label{nufull}
\ee
The magnetic field on the equator of magnetar is related to the dipole moment as 
\be
 B_s =  \frac{ \mu_0}{4 \pi} \frac{ \mu} {r_s^3 }. \label{bsandmu}
\ee
where $r_s$ is the radius of magnetar.
Then we can write Eq. (\ref{nufull}) in terms of surface magnetic field
\be
\nu^2 =  \frac {G } {c^4} \left ( \frac{ \mu_0}{4 \pi} \right )^{-1}  B_s^2 r_s^2  \left (  \frac{r_s} { r_{\rm Sch} } \right)^4 ,  \label{nufull2}
\ee
where $r_{\rm Sch }= 2GM / c^2$ is the Schwarzschild radius. 
Inserting $B_s = 10^{11} {\rm T}$ and $r_s = r_{\rm ps} = 1.5 r_{\rm Sch}$, we have $\nu^2 = 1.6 \times 10^{-8}$. For this value of $\nu^2$, 
\be
{\bar a} = {\bar a}_{\rm Sch}  -4.9 \times 10^{-8}  ,   ~~{\bar b} = {\bar b}_{\rm Sch} + 2.7 \times 10^{-6} \label{abbarem}
\ee
where $ {\bar a}_{\rm Sch} = 1$ and  ${\bar b}_{\rm Sch} = -0.40$ are the coefficients for the Schwarzschild black hole. 
We conclude that the magnetic correction to the deflection angle in the Einstein-Maxwell gravity is too small to give any observable consequence. 

\section{Deflection Angle in Einstein-Born-Infeld gravity}

When a light ray is passing close to a compact object with very strong electric or magnetic field, the nonlinear elctromagnetic effects can be significant. 
In the general relativistic theory coupled with linear electrodynamics, the electromagnetic wave and the gravitational wave follow the same geodesic made by mass, charge, and angular momentum. However, when it is coupled with nonlinear electrodynamics, the null geodesic of the electromagnetic wave is different from the null geodesic of the gravitational wave due to the nonlinear coupling with the background electromagnetic field. 
So we need to investigate the effective metric made by a compact object with mass and magnetic dipole. 

We consider the Einstein-Born-Infeld action described by \cite{Born,BornInfeld}
\be 
S = \int d^4 x \sqrt{-g} \left ( \frac{R}{16 \pi } +  {\cal L} \right ) , \label{EBIaction}
\ee
\be
 {\cal L} = \beta^2 \left ( 1- \sqrt{ 1 + \frac{2 S}{\beta^2} - \frac{P^2}{\beta^4} } \right ) ,
 \label{cbilagran}
\ee
where $\beta$ is the classical Born-Infeld parameter characterizing the possible maximum value of the field strength, $S$ and $P$ are Lorentz-invariants defined by
\be
 S = \frac{1}{4} F_{\mu \nu} F^{\mu \nu} = \frac{1}{2} ( {\bf B}^2 -  {\bf E}^2 )  , 
 ~~~ P = \frac{1}{4} F_{\mu \nu} {\tilde F}^{\mu \nu} =  {\bf E} \cdot {\bf B} ,
 \label{defSP}
\ee
$F_{\mu \nu} = \partial_\mu A_\nu - \partial_\nu A_\mu $ is the field strength tensor,  
and ${\tilde F}_{\mu \nu}  = \frac{1}{2} \epsilon_{\mu \nu \alpha \beta} F^{\alpha \beta} $ is the dual tensor. 
The effective metric for photon can be obtained from \cite {Plebanski,Novello,Delorenci,Breton,Eiroa}
\be 
 {\tilde g}^{\mu\nu} = g^{\mu\nu} + \frac {{\cal L}_{SS}} {{\cal L}_S} F^{\mu \alpha} F_\alpha^{~\nu} , \label{g_eff}
\ee
where $g^{\mu\nu}$ is the metric function of the gravitational wave, ${\cal L}_S = \partial {\cal L} / \partial S$, and ${\cal L}_{SS} = \partial^2 {\cal L} / \partial S^2$. 

Although the electromagnetic field is strong enough to consider the nonlinear effect, the metric is still dominated by mass near the photon sphere. Thus we can obtain the effective metric as a perturbation of the Schwarzschild metric. 
To the leading order, we can approximate
\be 
\frac {{\cal L}_{SS}} {{\cal L}_S} = - \frac{1}{\beta^2} \left ( 1 + \frac{2S} {\beta^2}  \right )^{-1} \rightarrow -  \frac{1} { \beta^2},
\ee
and obtain the effective metric from 
\be  
{\tilde g}^{\mu\nu} = g^{\mu\nu} - \frac{1}{\beta^2}  F^{\mu \alpha} F^{\nu}_{~\alpha} \equiv  g^{\mu\nu} + \delta  g^{\mu\nu} . \label{gtilde}
\ee

Substituting the electromagnetic field tensor from the four-potential in Eq. (\ref{four-pot})
\be
F_{\mu \nu} = \partial_\mu A_\nu -\partial_\nu A_\mu 
 = \begin{pmatrix} 0  & 0 &0  & 0 \cr 
0 & 0  &  0 & -\psi_r \cr  
0&  0& 0 &-\psi_z  \cr
0&  \psi_r& \psi_z & 0  \end{pmatrix} ,  \label{Fco}
\ee
where the subscripts $r$ and $z$ denote the partial derivatives, the leading-order nonzero components of $ \delta  g^{\mu\nu} $ are
\bea 
\delta  g^{11} &=& \frac{1}{\beta^2} \frac{\psi_r^2}{r^2}    \nonumber  \\
\delta  g^{12} &=& \delta  g^{21} = \frac{1}{\beta^2} \frac{\psi_r \psi_z}{r^2}    \nonumber  \\
\delta  g^{22} &=& \frac{1}{\beta^2} \frac{\psi_z^2}{r^2}    \nonumber  \\
\delta  g^{33} &=& \frac{1}{\beta^2} \frac{\psi_r \psi_z}{r^2}    \label{delgmunu}
\eea
In general one can find ${\tilde g}_{\mu\nu}$ from the inverse matrix of  (\ref{gtilde}). Using the explicit forms of metric and four-potential given by Eqs.  (\ref{erho})-(\ref{psi}), we can find the effective metric. 
On the equatorial plane, $\psi_r = - {\mu}/{r^2}, \psi_z = 0$, 
we obtain the leading order effective metric as 
\be 
ds^2 = \left ( 1 - \frac{ 2GM}{r} \right ) dt^2 -   \left( 1 + \frac{ 2GM} {r} + \frac {\mu^2}{\beta^2 r^6} \right ) dr^2 - r^2 \left ( 1 + \frac{2 GM } {r} +  \frac {\mu^2}{\beta^2 r^6}   \right ) d \phi^2 .
\ee
where we have neglected the Maxwell terms because their contribution is negligible as shown in the previous section.

Now we compute the deflection angle by defining 
\be 
x = r \sqrt{1 +  \frac{ 2GM} {r} + \frac {\mu^2}{\beta^2 r^6 } } = r \left (1 +  \frac{ GM} {r} + \frac {\mu^2}{2 \beta^2 r^6 } + \cdots \right )  , 
\ee
or 
\be 
r = x  \left ( 1 -  \frac{ GM} {x} - \frac {\mu^2}{2 \beta^2 x^6 }+\cdots \right ) .
\ee
In terms of $x$, we can write the effective metric in the standard form, Eq. (\ref{standardformABC}), where
\be
 A(x) = 1 - \frac{ 2GM}{x}  , ~ B(x) =  1 + \frac{ 2GM} {x} + \frac {\mu^2}{\beta^2 x^6} , ~  C(x) =  x^2 .  \label{metricABCBI}
\ee
Setting $x_s = 2GM=1$,  we repeat the computation to find the deflection angle with
\be
A(x) = 1 - \frac{1}{x}  , ~ B(x) =  1 + \frac{1} {x} + \frac {\gamma^2}{ x^6}  , ~  C(x) =  x^2 ,  \label{scaledABCBI}
\ee
where 
\be
\gamma^2 =  {\mu^2}/{\beta^2 (2GM)^6} .  \label{defgamma}
\ee  
The crucial difference from the Einstein-Maxwell case is  
\be 
R(z, x_0 )= 2 \left [ 1 + \frac{ \gamma^2} { 2}   \frac{1} { 1 - \frac{1-z}{x_0}} \frac { (1-z)^6} { x_0^6} + \cdots \right ] .  \label{rzx0leadingBI}
\ee

To the leading order, we found $\bar a$ and $\bar b$ in Eqs. (\ref{abar}) and (\ref{bbar}) as
\be
\bar a =  1 + \left ( \frac{2}{3} \right )^5 \gamma^2 , \label{abarBI}
\ee
\be 
b_R = b_{R,{\rm Sch}} + c^{\prime}_2 \left ( \frac{2}{3} \right )^6 \gamma^2 ,   \label{bRfinalBI}
\ee
where \ $c^{\prime}_2$ is given by the integral
\be 
c^{\prime}_2 = \int_0^1 3 \left [ \frac{ (1-z)^6}{1 +2 z } - 1 \right ] \frac{dz } {\sqrt{ z^2 - \frac{2}{3} z^3 }} = -9.3540. 
\ee 
Inserting Eqs. (\ref{abarBI}) and (\ref{bRfinalBI}) in Eq. (\ref{bbar}), we have
\be 
\bar b = - \pi + 2 \ln [ 6 (2 - \sqrt{3}) ] + \ln 6 + \gamma^2 \left [ c^{\prime}_2 \left (  \frac{2}{3} \right )^6 + \left (  \frac{2}{3} \right )^5 \ln6  \right ] = {\bar b}_{\rm Sch} - 0.5852 \gamma^2 .
\ee
Finally we express the deflection angle as
\be 
\alpha (\theta ) = - \left [1 + \left( \frac{2}{3} \right )^5 \gamma^2  \right ] \ln \left ( \frac{ \theta D_{OL} } { u_m } -1 \right ) + \ln [216 ( 7 - 4 \sqrt{3} ) ]  - \pi - 0.5852 \gamma^2 . 
\ee
Contrary to the result in the Einstein-Maxwell theory, the strong deflection coefficient $\bar a$ is larger and $\bar b$ is smaller than their Schwarzschild counterparts. Also the overall effect of the magnetic dipole is increasing the deflection angle. 

Now we also consider the numerical estimation for the magnetar considered before. 
Restoring the simplified units in Eq. (\ref{defgamma}), we have
\be
\gamma^2 =  \frac {1}{\beta^2} \left ( \frac{ \mu_0}{4 \pi} \frac{ \mu}{ \left ( {2 GM} /{c^2} \right )^3 } \right )^2   .  \label{gammafull}
\ee
Using the relation given by Eq. (\ref{bsandmu}), we can write the above equation in terms of the surface magnetic field as
\be
\gamma^2 =  \frac {B_s^2}{\beta^2} \left ( \frac { r_s} {2GM/ c^2}  \right )^6  .  \label{gammabsrs}
\ee
For the same mass $M= 1.4 M_S$  and surface magnetic field $B_s = 10^{11} {\rm T}$, and assuming that the Born-Infeld parameter is one order of magnitude larger than $B_s$, $\beta = 10^{12} \rm T$, we have $\gamma^2 = 0.11$. For this value of $\gamma^2$, 
\be
{\bar a} = {\bar a}_{\rm Sch}  + 0.015,   ~~{\bar b} = {\bar b}_{\rm Sch}  -0.067,  
\ee
where $ {\bar a}_{\rm Sch} = 1$ and  ${\bar b}_{\rm Sch} = -0.40$. 
Compared with Eq. (\ref{abbarem}), the contribution of magnetic dipole to deflection angle might not be negligible and lead to observable consequences depending on the Born-Infeld parameter. 

\section{Conclusion}
We consider the trajectory of light near the photon sphere of a compact object with strong magnetic field. We compute the deflection angle on the equatorial plane in both the Einstein-Maxwell gravity and the Einstein-Born-Infeld gravity.
In the Einstein-Maxwell gravity, we use the asymptotic metric distorted by magnetic dipole moment. 
We use the effective metric accomodating the nonlinear electrodynamic effect in the Einstein-Born-Infeld gravity.
The deflection angle are represented as perturbations from the Schwarzschild case. 
We applied the result to a magnetar to estimate the strong deflection coefficients numerically. The contribution of magnetic dipole to the deflection angle is negligibly small in the Einstein-Maxwell gravity. However, in the Einstein-Born-Infeld gravity, the contribution might not be negligible.

When we compute the deflection angle in the Einstein-Born-Infeld gravity, we assume the magnetic field near the photon sphere is below the Born-Infeld parameter. 
Because the Born-Infeld parameter $\beta$ is the possible maximum field strength, one might consider the case where the magnetic field $B$ near the photon sphere is saturated to $\beta$. However, when $B$ has a constant  value $\beta$, no metric correction can be caused by nonlinear electromagnetic effect. 
In general relativity the bending of light is caused by the gradient of the metric. So no deflection of light can be caused by uniform field.

For a numerical estimation, we consider a magnetar compact enough to have a photon sphere. It is known that black holes might have very weak magnetic field \cite{EHT2021}. Similar numerical estimations can be considered assuming black holes with strong magnetic field might exist.
In this study we compute the bending angle for a special case where a light ray is passing on the equatorial plane of the magnetic dipole. It will be interesting to study the bending angle for an arbitrary orientation of the magnetic dipole.

\section*{Acknowledgements}
We would like to thank M. I. Park and C. Ho for discussion and help.

\appendix

\end{document}